\documentclass[12pt]{article}
\usepackage{amsmath}

\begin{document}

\author{Jos\'{e} L. Cereceda \\
\textit{C/Alto del Le\'{o}n 8, 4A, 28038 Madrid, Spain}}
\title{\textbf{A simple proof of the converse of \\
\vspace{-2.mm} Hardy's theorem}}
\date{\today}

\maketitle

\begin{abstract}
In this paper we provide a simple proof of the fact that for a system of two
spin-$\frac{1}{2}$ particles, and for a choice of a set of observables,
there is a unique state which shows Hardy-type nonlocality. Moreover, an
explicit expression for the probability that an ensemble of particle pairs
prepared in such a state exhibits a Hardy-type nonlocality contradiction
is given in terms of two independent parameters related to the observables
involved. Incidentally, a wrong statement expressed in Mermin's proof of the
converse [N.D.~Mermin, Am. J. Phys. \textbf{62}, 880 (1994)] is pointed out.

\vspace{.2cm}

\end{abstract}

\section{Introduction}

In his pioneering paper [1], John Bell established the first of a series of
inequalities now collectively known as ``Bell inequalities'' which
demonstrate the incompatibility between the predictions of quantum theory
and the limitations imposed by a local and realistic world view. Bell showed
that for an Einstein-Podolsky-Rosen-Bohm gedanken experiment [2,3], the
quantum-mechanical predictions for an ensemble of pairs of spin-$\frac{1}{2}$
particles prepared in the singlet state violate an inequality which must
necessarily be satisfied by any physical theory based on the reality and
locality postulates. More recently Hardy [4], by means of an ingenious
thought experiment, managed to find a proof of nonlocality for two particles
without using inequalities, with each of the particles living in an
effective two-dimensional Hilbert state space. A more systematic version of
this proof was given by Hardy himself [5], and then by Goldstein [6], Jordan
[7,8], Aravind [9], and Cereceda [10], among other authors. In its simplest
form, Hardy's nonlocality theorem involves the values of four dichotomic
observables (two for each particle), and states that a direct contradiction
between the quantum-mechanical and locally realistic predictions concerning
such values can arise for some probabilistic fraction of an ensemble of particle
pairs configured in an entangled state, provided that the entanglement is not
maximal.

The converse of this result has also been proved [7,11-13]: for any choice
of two different observables for each particle, a state can be found which
admits a Hardy-type nonlocality contradiction. However, while Jordan's
valuable treatment of the converse lacks some of the clearness required by
a beginner student (in fact, Jordan's proof can be greatly simplified, as shown
in Refs.\ [11-13]), Kar's remarkable proof deals only with abstract projection
operators, without any reference to concrete physical measurable quantities.
Regarding Mermin's instructive proof (see Sections I and II of Ref.\ [13]), I
maintain that his conclusion that, ``{\ldots}we can do the trick with any two
choices of nontrivial local one-particle observables to be measured in mode 1
on the right and on the left, and any state $\left| \Psi \right\rangle $ of the
form (2) with three nonzero amplitudes $\alpha $, $\beta $, and $\gamma $'',
is not entirely correctly stated. Indeed, from Eq.\ (7) of Ref.\ [13], we can
see that the magnitude of $\gamma $ is fixed by the quantities $\left|
\left\langle 2G\mid 1R\right\rangle _{r}\right| $, $\left| \left\langle
2G\mid 1G\right\rangle _{r}\right| $, $\left| \left\langle 2G\mid
1R\right\rangle _{l}\right| $, and $\left| \left\langle 2G\mid
1G\right\rangle _{l}\right| $. Likewise, from Eqs.\ (3), (4), and (7) of Ref.\
[13], we can readily express the magnitude of $\alpha $ and $\beta $ in terms
of the four quantities above. On the other hand, from Eq.\ (3) [(4)] of Ref.\
[13], it follows that the phase of $\beta $ [$\alpha $] is in turn determined
by the phases of $\left\langle 2G\mid 1R\right\rangle _{r}$ and $%
\left\langle 2G\mid 1G\right\rangle _{r}$ [$\left\langle 2G\mid
1R\right\rangle _{l}$ and $\left\langle 2G\mid 1G\right\rangle _{l}$]
(without loss of generality, we may choose the amplitude $\gamma $ to be
real). So, strictly speaking, the trick will only work for that state $%
\left| \Psi \right\rangle $ whose (nonzero) coefficients $\alpha $, $\beta $%
, and $\gamma $ \textit{do }satisfy simultaneously the above set of Eqs.\
(3), (4), and (7). What Mermin actually proves is a version of the converse
of Hardy's theorem, in which both nontrivial one-particle observables for
mode 1 on the right and on the left are \textit{quite} arbitrary (in
Mermin's words, ``You can pick any two one-particle observables you like for
the detectors to measure when their switches are set to 1''), both
one-particle observables for mode 2 on the right and on the left are \textit{%
almost} arbitrary (in Mermin's words, ``The choices of the two observables
measured by the detectors in modes 2 are almost as flexible''), and where,
as a result, the state $\left| \Psi \right\rangle $ turns out to be fixed
(up to an arbitrary overall phase factor) by the above choice of four
observables through the Eqs.\ (2), (3), (4), and (7) of Ref.\ [13]. As Mermin
then shows, the probability of getting the ``impossible'' $22GG$ events is
maximum for those choices of observables for which $\left| \left\langle 2G\mid
1G\right\rangle _{l}\right| ^{2}$ $=\left| \left\langle 2G\mid
1G\right\rangle _{r}\right| ^{2}=\tau ^{-1}$, and $\left| \left\langle
2G\mid 1R\right\rangle _{l}\right| ^{2}=\left| \left\langle 2G\mid
1R\right\rangle _{r}\right| ^{2}=1-\tau ^{-1}$, with $\tau $ being the
golden mean, $\frac{1}{2}(1+\sqrt{5})$.

Following Mermin's treatment, in this paper we present a direct,
easy-to-grasp proof of the converse of Hardy's theorem for two spin-$\frac{1%
}{2}$ particles (which we call $a$ and $b$) that achieves a certain
generality while retaining the mathematical simplicity. This proof makes clear
the fact that, for two spin-$\frac{1}{2}$ particles and for a given choice of
observables, there is a unique state which satisfies Hardy's nonlocality. Moreover,
a graphical representation of the probability for a Hardy-type contradiction
(see Eq.\ (9) and Fig.\ 1 below) is given in terms of two independent parameters
related to the observables involved. For this purpose, we consider the spin
components of each particle along directions lying within the \textit{x-z}
plane, so that we shall deal with the observables $S(\theta _{a})$ and $%
S(\theta _{a}^{^{\prime }})$ for particle $a$, and $S(\theta _{b})$ and $%
S(\theta _{b}^{^{\prime }})$ for particle $b$, with $S(\theta _{a})$, $%
S(\theta _{a}^{^{\prime }})$, $S(\theta _{b})$, and $S(\theta _{b}^{^{\prime
}})$ being spin components along directions making an angle $\theta _{a}$, $%
\theta _{a}^{^{\prime }}$, $\theta _{b}$, and $\theta _{b}^{^{\prime }}$,
respectively, with the \textit{z} axis in the \textit{x-z} plane (of course
it is understood that the \textit{z} axis is defined with reference to in
general different coordinate systems for each of the particles). Further, as
usual, it is assumed that each of the spin components yields the outcome $+1$
or $-1$ (in appropriate units) if measured.

\section{The converse of Hardy's theorem}

A two-particle state $\left| \eta \right\rangle $ exhibits a Hardy-type
nonlocality contradiction if there exist two observables for each
particle---say, $S(\theta _{a})$ and $S(\theta _{a}^{^{\prime }})$ for
particle $a$, and $S(\theta _{b})$ and $S(\theta _{b}^{^{\prime }})$ for
particle $b$---such that the following four conditions are simultaneously
fulfilled for the state $\left| \eta \right\rangle ^{1}$
\begin{align}
P_{\eta }(S(\theta _{a})=+1,S(\theta _{b})=+1) &=0,  \tag{1a} \\
P_{\eta }(S(\theta _{a})=-1,S(\theta _{b}^{^{\prime }})=-1) &=0,  \tag{1b}
\\
P_{\eta }(S(\theta _{a}^{^{\prime }})=-1,S(\theta _{b})=-1) &=0,  \tag{1c}
\\
P_{\eta }(S(\theta _{a}^{^{\prime }})=-1,S(\theta _{b}^{^{\prime }})=-1)
&>0,  \tag{1d}
\setcounter{equation}{1}
\end{align}
where, for example, $P_{\eta }(S(\theta _{a})=+1,S(\theta _{b})=+1)$ denotes
the probability that a joint measurement of the observables $S(\theta _{a})$
and $S(\theta _{b})$ on particles $a$ and $b$, respectively, gives the
result $+1$ for both observables when the particles are described by the
state vector $\left| \eta \right\rangle $. As we have said, every entangled
state except those maximally entangled (such as the singlet state considered
by Bell) shows Hardy-type contradiction, that is, there are observables $%
S(\theta _{a})$, $S(\theta _{a}^{^{\prime }})$, $S(\theta _{b})$ and $%
S(\theta _{b}^{^{\prime }})$ (in fact, there are an infinite number of them
[6,7,10]) for which the conditions (1a)-(1d) are fulfilled for the entangled
state. That conditions (1a)-(1d) lead to a contradiction with the assumption
of local realism can be seen as follows [5]. Consider a particular run of
the experiment for which the observables $S(\theta _{a}^{^{\prime }})$ and $%
S(\theta _{b}^{^{\prime }})$ are measured on a pair of spacelike separated
particles $a$ and $b$ in the state $\left| \eta \right\rangle $, and the
results $S(\theta _{a}^{^{\prime }})=-1$ and $S(\theta _{b}^{^{\prime }})=-1$
are obtained. From Eq.\ (1d), there is a nonzero probability for these joint
measurement results to occur. By invoking local realism, and taking into
account the constraint in Eq.\ (1c), one could assert that a result $S(%
\theta _{b})=+1$ would have been obtained if, instead of $S(\theta _{b}^
{^{\prime }})$, the observable $S(\theta _{b})$ had been measured on particle
$b$. Similarly, from Eq.\ (1b), and applying local realism, one might conclude
that a result $S(\theta _{a})=+1$ would have been obtained for a measurement
of $S(\theta _{a})$ on particle $a$. In this way, the quantum predictions (1d),
(1c), and (1b), together with the assumption of local realism, allows one to
deduce that there must be a nonzero probability to obtain the results $S(%
\theta _{a})=+1$ and $S(\theta _{b})=+1$ in a joint measurement of the observables
$S(\theta _{a})$ and $S(\theta _{b})$. However, from Eq.\ (1a), we cannot have
simultaneously the results $S(\theta _{a})=+1$ and $S(\theta _{b})=+1$ for any
pairs of particles described by $\left| \eta \right\rangle $. Hence a
contradiction between quantum mechanics and local realism arises without using
inequalities.

It is convenient to introduce the relative angles $\theta _{1}=\theta
_{a}^{^{\prime }}-\theta _{a}$ and $\theta _{2}=\theta _{b}^{^{\prime
}}-~\!\theta _{b}$. We are now ready to prove the converse of Hardy's theorem.
Specifically, we show that for any choice of $\theta _{a}$, $\theta
_{a}^{^{\prime }}$, $\theta _{b}$, and $\theta _{b}^{^{\prime }}$ (or,
equivalently, for any choice of $\theta _{a}$, $\theta _{b}$, $\theta _{1}$,
and $\theta _{2}$), there exists a two-particle state $\left| \eta
\right\rangle $ satisfying the above conditions (1a)-(1d), provided that
both $\theta _{1}$ and $\theta _{2}$ are not an integral multiple of $\pi $.
To do this, we first write the quantum state in terms of eigenvectors for
the $S(\theta _{a})$ component of particle $a$ and $S(\theta _{b})$
component of particle $b$
\begin{multline}
\left| \eta \right\rangle
= c_{++}\left| S(\theta _{a})=+1\right\rangle \left| S(\theta
_{b})=+1\right\rangle +c_{+-}\left| S(\theta _{a})=+1\right\rangle \left|
S(\theta _{b})=-1\right\rangle   \\
+ c_{-+}\left| S(\theta _{a})=-1\right\rangle
\left| S(\theta _{b})=+1\right\rangle +c_{--}\left| S(\theta
_{a})=-1\right\rangle \left| S(\theta _{b})=-1\right\rangle ,
\end{multline}
where, for example, $\left| S(\theta _{a})=+1\right\rangle $ represents a
state of spin-up for particle $a$ along a direction inclined at an angle $%
\theta _{a}$ to the \textit{z} axis in the \textit{x-z} plane. The
coefficients $c_{++}$, $c_{+-}$, $c_{-+}$, and $c_{--}$ (which, for
simplicity, are assumed to be real---the extension to complex coefficients
offering no difficulty) obey the normalization condition 
\begin{equation}
c_{++}^{2}+c_{+-}^{2}+c_{-+}^{2}+c_{--}^{2}=1.
\end{equation}
Of course the expansion (2) is always possible since the set of product
vectors $\left\{ \left| S(\theta _{a})=i\right\rangle \left| S(\theta
_{b})=j\right\rangle \right\} $, with $i,j=+1$ or $-1$, span the total
four-dimensional Hilbert space associated with the spin of particles $a$ and 
$b$. Clearly, condition (1a) does hold for the state vector $\left| \eta
\right\rangle $ if, and only if, $c_{++}=0$. On the other hand, by
expressing the eigenvectors $\left| S(\theta _{a})=+1\right\rangle $ and $%
\left| S(\theta _{a})=-1\right\rangle $ in terms of $\left| S(\theta
_{a}^{^{\prime }})=+1\right\rangle $ and $\left| S(\theta _{a}^{^{\prime
}})=-1\right\rangle $, 
\begin{align}
\left| S(\theta _{a})=+1\right\rangle &= \cos \frac{\theta _{1}}{2}\left|
S(\theta _{a}^{^{\prime }})=+1\right\rangle -\sin \frac{\theta _{1}}{2}%
\left| S(\theta _{a}^{^{\prime }})=-1\right\rangle ,  \tag{4a} \\
\left| S(\theta _{a})=-1\right\rangle &= \sin \frac{\theta _{1}}{2}\left|
S(\theta _{a}^{^{\prime }})=+1\right\rangle +\cos \frac{\theta _{1}}{2}%
\left| S(\theta _{a}^{^{\prime }})=-1\right\rangle ,  \tag{4b}
\end{align}
and the eigenvectors $\left| S(\theta _{b})=+1\right\rangle $ and $\left|
S(\theta _{b})=-1\right\rangle $ in terms of $\left| S(\theta _{b}^{^{\prime
}})=+1\right\rangle $ and $\left| S(\theta _{b}^{^{\prime
}})=-1\right\rangle $, 
\begin{align}
\left| S(\theta _{b})=+1\right\rangle &= \cos \frac{\theta _{2}}{2}\left|
S(\theta _{b}^{^{\prime }})=+1\right\rangle -\sin \frac{\theta _{2}}{2}%
\left| S(\theta _{b}^{^{\prime }})=-1\right\rangle ,  \tag{5a} \\
\left| S(\theta _{b})=-1\right\rangle &= \sin \frac{\theta _{2}}{2}\left|
S(\theta _{b}^{^{\prime }})=+1\right\rangle +\cos \frac{\theta _{2}}{2}%
\left| S(\theta _{b}^{^{\prime }})=-1\right\rangle ,  \tag{5b}
\setcounter{equation}{5}
\end{align}
it is immediate to see that the fulfillment of conditions (1b) and (1c) is
equivalent to requiring, respectively, that 
\begin{equation}
c_{--}\cos \left( \frac{\theta _{2}}{2}\right) -c_{-+}\sin \left( \frac{%
\theta _{2}}{2}\right) =0,  
\end{equation}
and 
\begin{equation}
c_{--}\cos \left( \frac{\theta _{1}}{2}\right) -c_{+-}\sin \left( \frac{%
\theta _{1}}{2}\right) =0. 
\end{equation}
(Note that the pair of Eqs.\ (6) and (7) is the equivalent of Mermin's pair
of Eqs.\ (3) and (4) in Ref.\ [13].) Relations (6) and (7), together with the
normalization condition (3), allow us to solve for the coefficients $c_{-+}$%
, $c_{+-}$, and $c_{--}$, expressed as a function of the parameters $\theta
_{1}$ and $\theta _{2}$. A straightforward calculation gives 
\begin{align}
c_{-+} &= \pm \frac{\tan \left( \frac{\theta _{1}}{2}\right) }{\left[ \tan
^{2}\left( \frac{\theta _{1}}{2}\right) +\tan ^{2}\left( \frac{\theta _{2}}{2%
}\right) +\tan ^{2}\left( \frac{\theta _{1}}{2}\right) \tan ^{2}\left( \frac{%
\theta _{2}}{2}\right) \right] ^{\frac{1}{2}}}, \tag{8a} \\[2.mm]
c_{+-} &= \pm \frac{\tan \left( \frac{\theta _{2}}{2}\right) }{\left[ \tan
^{2}\left( \frac{\theta _{1}}{2}\right) +\tan ^{2}\left( \frac{\theta _{2}}{2%
}\right) +\tan ^{2}\left( \frac{\theta _{1}}{2}\right) \tan ^{2}\left( \frac{%
\theta _{2}}{2}\right) \right] ^{\frac{1}{2}}}, \tag{8b} \\[2.mm]
c_{--} &= \pm \frac{\tan \left( \frac{\theta _{1}}{2}\right) \tan \left( 
\frac{\theta _{2}}{2}\right) }{\left[ \tan ^{2}\left( \frac{\theta _{1}}{2}%
\right) +\tan ^{2}\left( \frac{\theta _{2}}{2}\right) +\tan ^{2}\left( \frac{%
\theta _{1}}{2}\right) \tan ^{2}\left( \frac{\theta _{2}}{2}\right) \right]
^{\frac{1}{2}}}.  \tag{8c} 
\setcounter{equation}{8}
\end{align}
The state vector (2) with coefficients $c_{-+}$, $c_{+-}$, and $c_{--}$
given by Eqs.\ (8a)-(8c), and the coefficient $c_{++}$ set to zero, satisfies
the Hardy equations (1a)-(1c). It remains to check out that, for such a state,
the probability $P_{\eta }(S(\theta _{a}^{^{\prime }})=-1,S(\theta
_{b}^{^{\prime }})=-1)$ is nonzero almost everywhere. This probability is a
measure of the statistical fraction of particle pairs prepared in the state $%
\left| \eta \right\rangle $, for which a joint measurement of $S(\theta
_{a}^{^{\prime }})$ and $S(\theta _{b}^{^{\prime }})$ gives results which,
when combined with the quantum predictions (1a)-(1c), cannot be explained by
any locally realistic theory. On using the coefficients (8a)-(8c) in
expansion (2), and replacing the eigenvectors $\left| S(\theta
_{a})=+1\right\rangle $, $\left| S(\theta _{a})=-1\right\rangle $, $\left|
S(\theta _{b})=+1\right\rangle $, and $\left| S(\theta _{b})=-1\right\rangle 
$ with the corresponding expression (4a), (4b), (5a), and (5b), one finds
that the above probability is given by 
\begin{align}
P_{\eta }(S(\theta _{a}+\theta _{1})\, &= -1,S(\theta _{b}+\theta _{2})=-1) 
\nonumber \\
&= \frac{\sin ^{2}\left( \frac{\theta _{1}}{2}\right) \sin ^{2}\left( \frac{%
\theta _{2}}{2}\right) }{\tan ^{2}\left( \frac{\theta _{1}}{2}\right) +\tan
^{2}\left( \frac{\theta _{2}}{2}\right) +\tan ^{2}\left( \frac{\theta _{1}}{2%
}\right) \tan ^{2}\left( \frac{\theta _{2}}{2}\right) }.  
\end{align}
This function is represented in Fig.\ 1 for the ranges of variation $0^{\circ
}\leq \theta _{1}\leq 360^{\circ }$ and $0^{\circ }\leq \theta _{2}\leq
360^{\circ }$. The probability function (9) is found to vanish wherever $%
\theta _{1}=n_{1}\pi $, or $\theta _{2}=n_{2}\pi $, with $n_{1},n_{2}=0,\pm
1,\pm 2,\ldots $ . Except for these values of $\theta _{1}$ or $\theta _{2}$
the above probability is positive. The vanishing of (9) corresponds to the
case in which the observables $S(\theta _{a})$ and $S(\theta _{a}^{^{\prime
}})$, or $S(\theta _{b})$ and $S(\theta _{b}^{^{\prime }})$, happen to
commute. So we have proved that for any two pairs of noncommuting
observables, $S(\theta _{a})$ and $S(\theta _{a}^{^{\prime }})$ for particle 
$a$, and $S(\theta _{b})$ and $S(\theta _{b}^{^{\prime }})$ for particle $b$%
, a two-particle state exists that fulfills the conditions (1a)-(1d). As we
have said, this state is given by Eq.\ (2) with $c_{++}=0$, and $c_{-+}$, $%
c_{+-}$, and $c_{--}$ given by Eqs.\ (8a)-(8c). It is important to notice
that the state vector (2) satisfying the set of equations (1a)-(1d) for
given $\theta _{a}$, $\theta _{a}^{^{\prime }}$, $\theta _{b}$, and $\theta
_{b}^{^{\prime }}$ is unique up to an arbitrary overall phase factor. This
follows from the fact that, (i) the eigenvectors $\left| S(\theta
_{a})=+1\right\rangle $ and $\left| S(\theta _{a})=-1\right\rangle $ [ $%
\left| S(\theta _{b})=+1\right\rangle $ and $\left| S(\theta
_{b})=-1\right\rangle $] appearing in expansion (2) are uniquely determined
by the angle $\theta _{a}$ [$\theta _{b}$]; and (ii) the expansion
coefficients $c_{-+}$, $c_{+-}$, and $c_{--}$ satisfying the relations (3),
(6), and (7) (with the coefficient $c_{++}$ set to zero in Eq.\ (3)) are
unique (up to a common sign factor) for fixed values of $\theta _{1}$ and $%
\theta _{2}$. This uniqueness implies that no mixture state for two spin-$%
\frac{1}{2}$ particles admits Hardy's nonlocality for fixed choice of
observables [11,12].

Now, according to Mermin's theory [13], and making the identifications 
\begin{eqnarray}
\left| S(\theta _{a}^{^{\prime }})=-1\right\rangle \equiv \left|
2G\right\rangle _{l}; &&\left| S(\theta _{b}^{^{\prime }})=-1\right\rangle
\equiv \left| 2G\right\rangle _{r},  \nonumber \\
\left| S(\theta _{a}^{^{\prime }})=+1\right\rangle \equiv \left|
2R\right\rangle _{l}; &&\left| S(\theta _{b}^{^{\prime }})=+1\right\rangle
\equiv \left| 2R\right\rangle _{r},  \nonumber \\
\left| S(\theta _{a})=-1\right\rangle \equiv \left| 1G\right\rangle _{l};
&&\left| S(\theta _{b})=-1\right\rangle \equiv \left| 1G\right\rangle _{r}, 
\nonumber \\
\left| S(\theta _{a})=+1\right\rangle \equiv \left| 1R\right\rangle _{l};
&&\left| S(\theta _{b})=+1\right\rangle \equiv \left| 1R\right\rangle _{r}, 
\end{eqnarray}
it follows that the probability (9) is maximum for those choices of $\theta
_{a}$, $\theta _{a}^{^{\prime }}$, $\theta _{b}$, and $\theta _{b}^{^{\prime
}}$ for which 
\begin{equation}
\left| \left\langle S(\theta _{a}^{^{\prime }})=-1\mid S(\theta
_{a})=-1\right\rangle \right| ^{2}=\left| \left\langle S(\theta
_{b}^{^{\prime }})=-1\mid S(\theta _{b})=-1\right\rangle \right| ^{2}=\tau
^{-1},  \tag{11a}
\end{equation}
and 
\begin{equation}
\left| \left\langle S(\theta _{a}^{^{\prime }})=-1\mid S(\theta
_{a})=+1\right\rangle \right| ^{2}=\left| \left\langle S(\theta
_{b}^{^{\prime }})=-1\mid S(\theta _{b})=+1\right\rangle \right| ^{2}=1-\tau
^{-1}.  \tag{11b}
\setcounter{equation}{11}
\end{equation}
From Eqs.\ (4b) and (5b), condition (11a) implies that $\cos ^{2}\left( \frac{%
\theta _{1}}{2}\right) =\cos ^{2}\left( \frac{\theta _{2}}{2}\right) =\tau
^{-1}$. Analogously, from Eqs.\ (4a) and (5a), condition (11b) implies that $%
\sin ^{2}\left( \frac{\theta _{1}}{2}\right) =\sin ^{2}\left( \frac{\theta
_{2}}{2}\right) =1-\tau ^{-1}$. For the ranges of variation $0^{\circ }\leq
\theta _{1}\leq 360^{\circ }$ and $0^{\circ }\leq \theta _{2}\leq 360^{\circ }$,
these conditions are fulfilled for the following pairs of values (see Fig.\
1): $(\theta _{1},\theta _{2})=(\theta _{0},\theta _{0}),(\theta _{0},2\pi
-\theta _{0}),(2\pi -\theta _{0},\theta _{0})$, and $(2\pi -\theta _{0},2\pi
-\theta _{0})$, with $\theta _{0}\simeq 76.3454^{\circ }$. For any one of
these pairs, the probability function (9) attains the maximum value $P_{\eta
}^{\text{max}}=1/\tau ^{5}$, while the squared coefficients $c_{-+}^{2}$, $%
c_{+-}^{2}$, and $c_{--}^{2}$ yielding this maximum probability are 
\begin{align}
c_{-+}^{2}= c_{+-}^{2} &= 1/\tau ^{2},  \nonumber \\
c_{--}^{2} &= 1/\tau ^{3}.  
\end{align}
Naturally, according to Eq.\ (3), we have $2/\tau ^{2}+1/\tau ^{3}=1$.

On the other hand, as mentioned above, the probability (9) is zero for
either $\theta _{1}=n_{1}\pi $ or $\theta _{2}=n_{2}\pi $, so that the
nonlocality argument will fail as soon as any one of the four quantities $%
\left| \left\langle S(\theta _{a}^{^{\prime }})=-1\mid S(\theta
_{a})=i\right\rangle \right| $, \linebreak $\left| \left\langle S(\theta _{b}
^{^{\prime%
}})=-1\mid S(\theta _{b})=j\right\rangle \right| $, with $i,j=+1$ or $-1$,
happens to vanish. This is just the reason why the choice of all four
observables $S(\theta _{a})$, $S(\theta _{a}^{^{\prime }})$, $S(\theta _{b})$%
, and $S(\theta _{b}^{^{\prime }})$, cannot be totally arbitrary. As Mermin
points out [13], in order for the argument to run, it is necessary that the
eigenvector $\left| S(\theta _{a}^{^{\prime }})=-1\right\rangle $ $\left[ 
\:\left| S(\theta _{b}^{^{\prime }})=-1\right\rangle \right] $ has
nonzero components along both $\left| S(\theta _{a})=+1\right\rangle $ and
\linebreak $%
\left| S(\theta _{a})=-1\right\rangle $ $[\:\left| S(\theta
_{b})=+1\right\rangle $ and $\left| S(\theta _{b})=-1\right\rangle $].
Clearly the fulfillment of this condition guarantees that the eigenvector $%
\left| S(\theta _{a}^{^{\prime }})=+1\right\rangle $ $\left[\:\left|
S(\theta _{b}^{^{\prime }})=+1\right\rangle \right] $ has equally nonzero
components along both $\left| S(\theta _{a})=+1\right\rangle $ and $\left|
S(\theta _{a})=-1\right\rangle $ $[\:\left| S(\theta _{b})=+1\right\rangle $
and $\left| S(\theta _{b})=-1\right\rangle $]. By the way we note that the
above conditions (11a) and (11b) defining the maxima of the probability
function (9) can be written equivalently as 
\begin{equation}
\left| \left\langle S(\theta _{a}^{^{\prime }})=+1\mid S(\theta
_{a})=-1\right\rangle \right| ^{2}=\left| \left\langle S(\theta
_{b}^{^{\prime }})=+1\mid S(\theta _{b})=-1\right\rangle \right| ^{2}=1-\tau
^{-1},  \tag{13a}
\end{equation}
and 
\begin{equation}
\left| \left\langle S(\theta _{a}^{^{\prime }})=+1\mid S(\theta
_{a})=+1\right\rangle \right| ^{2}=\left| \left\langle S(\theta
_{b}^{^{\prime }})=+1\mid S(\theta _{b})=+1\right\rangle \right| ^{2}=\tau
^{-1},  \tag{13b}
\setcounter{equation}{13}
\end{equation}
respectively. Likewise, the vanishing of any one of the four quantities
\linebreak $\left| \left\langle S(\theta _{a}^{^{\prime }})=+1\mid S(\theta
_{a})=i\right\rangle \right| $, $\left| \left\langle S(\theta _{b}^{^{\prime
}})=+1\mid S(\theta _{b})=j\right\rangle \right| $, means that the
probability (9) is zero.$^{2}$ It will further be noted that for the
particular case in which $\theta _{1}=\theta _{2}\equiv \theta $, expression
(9) reduces to 
\begin{equation}
P_{\eta }(S(\theta _{a}+\theta )=-1,S(\theta _{b}+\theta )=-1)=\frac{\sin
^{2}\theta }{4+4\sec ^{2}\left( \frac{\theta }{2}\right) },  
\end{equation}
while the coefficients $c_{-+}$, $c_{+-}$, and $c_{--}$ are found to be 
\begin{align}
c_{-+}=c_{+-}&= \pm \left[ 2+\tan ^{2}\left( \frac{\theta }{2}\right) \right]
^{-\frac{1}{2}},   \\
c_{--}&= \pm \left[ 1+2\cot ^{2}\left( \frac{\theta }{2}\right) \right] ^{-%
\frac{1}{2}}.  
\end{align}
The probability function (14) is plotted in Fig.\ 2.

\section{Concluding remarks}

We conclude by noting that the lack of one of the terms in expansion (2)
prevents the state vector $\left| \eta \right\rangle $ from being maximally
entangled. To see this, recall that, according to the Schmidt decomposition
theorem [9,14,15], any pure spin state for two spin-$\frac{1}{2}$ particles
can be expressed as a sum of two biorthogonal terms. In particular, the
state vector (2) with $c_{++}=0$, and $c_{-+}$, $c_{+-}$, and $c_{--}$ given
by Eqs.\ (8a)-(8c), can be put in the form
\pagebreak
\begin{equation}
\left| \eta \right\rangle =c_{+}\left| S(\phi _{a})=+1\right\rangle \left|
S(\phi _{b})=+1\right\rangle +c_{-}\left| S(\phi _{a})=-1\right\rangle
\left| S(\phi _{b})=-1\right\rangle ,  
\end{equation}
for some appropriate $c_{+}$ and $c_{-}$ depending on $\theta _{1}$ and $%
\theta _{2}$, and some appropriate $\phi _{a}$ $(\phi _{b})$ depending on $%
\theta _{a}$, $\theta _{1}$, and $\theta _{2}$ $(\theta _{b}$, $\theta _{1}$%
, and $\theta _{2})$, with the real coefficients $c_{+}$ and $c_{-}$
fulfilling $c_{+}^{2}+c_{-}^{2}=1$, and where the polar angles $\phi _{a}$
and $\phi _{b}$ define the orientation of two axes in the \textit{x-z}
plane. Without any loss of generality, we may assume that $c_{+}^{2}\geq
c_{-}^{2}$. The state in Eq.\ (17) is said to be maximally entangled if $%
c_{+} $ and $c_{-}$ are equal in absolute value, that is, if $%
c_{+}^{2}=c_{-}^{2}=\frac{1}{2}$. We want to show that the state vector (2)
with $c_{++}=0$ does not admit a decomposition like that of Eq.\ (17) for
which $c_{+}^{2}=c_{-}^{2}$. This follows in a rather straightforward way
from the easily confirmed fact that, whenever $\left| c_{+}\right| =\left|
c_{-}\right| $, the state vector (17) implies that 
\begin{equation}
P_{\eta }(S(\theta _{a})=+1,S(\theta _{b})=+1)=P_{\eta }(S(\theta
_{a})=-1,S(\theta _{b})=-1).  
\end{equation}
However, for the case in which $c_{++}$ is equal to zero, we have from Eq.\
(2) that $P_{\eta }(S(\theta _{a})=+1,S(\theta _{b})=+1)=0$. On the other
hand, the probability $P_{\eta }(S(\theta _{a})=-1,S(\theta
_{b})=-1)=c_{--}^{2}$ is in general different from zero, as can be seen from
Eq.\ (8c) or (16). This contradiction establishes that the state vector
satisfying the conditions (1a)-(1d) for given $\theta _{a}$, $\theta
_{a}^{^{\prime }}$, $\theta _{b}$, and $\theta _{b}^{^{\prime }}$, cannot be
maximally entangled. This is consistent of course with the fact that no
maximally entangled state for two spin-$\frac{1}{2}$ particles can give a
Hardy-type nonlocality contradiction [5-10].

For the sake of completeness, let us now briefly sketch how one can find the
decomposition (17) that obtains for a given state vector (2). (Recall that $%
\theta _{a}$, $\theta _{b}$, $\theta _{1}$, and $\theta _{2}$ (with $\theta
_{1}$, $\theta _{2}\neq 0$ or $\pi $) are arbitrary but fixed parameters,
and that the coefficient $c_{++}$ is set to zero throughout.) The density
matrix associated with the pure state (2) is $\rho =\left| \eta
\right\rangle \left\langle \eta \right| $, and the corresponding reduced
density matrices describing the spin of particles $a$ and $b$ are,
respectively, $\rho _{a}=\text{Tr}_{b}\,\rho $ and $\rho _{b}=\text{Tr}%
_{a}\,\rho $, where $\text{Tr}_{a}$ ($\text{Tr}_{b}$) denotes the trace
operation over the spin states of particle $a$ ($b$). In our case, $\rho $
is a $4\times 4$ symmetric matrix, whereas both $\rho _{a}$ and $\rho _{b}$
are $2\times 2$ symmetric matrices. All these density matrices are most
easily written down when the two orthonormal Hilbert space vectors $\left( 
\begin{smallmatrix}
1 \\ 
0
\end{smallmatrix}
\right) _{a}$ and $\left( 
\begin{smallmatrix}
0 \\ 
1
\end{smallmatrix}
\right) _{a}$ for particle $a$ are chosen such that
\begin{equation}
\left| S(\theta _{a})=+1\right\rangle \equiv \left( 
\begin{array}{l}
1 \\ 
0
\end{array}
\right) _{a},\;\;\;\;\left| S(\theta _{a})=-1\right\rangle \equiv \left( 
\begin{array}{l}
0 \\ 
1
\end{array}
\right) _{a},  \tag{19a}
\end{equation}
and, similarly, the basis vectors $\left( 
\begin{smallmatrix}
1 \\ 
0
\end{smallmatrix}
\right) _{b}$ and $\left( 
\begin{smallmatrix}
0 \\ 
1
\end{smallmatrix}
\right) _{b}$ for particle $b$ are chosen such that 
\begin{equation}
\left| S(\theta _{b})=+1\right\rangle \equiv \left( 
\begin{array}{l}
1 \\ 
0
\end{array}
\right) _{b},\;\;\;\;\left| S(\theta _{b})=-1\right\rangle \equiv \left( 
\begin{array}{l}
0 \\ 
1
\end{array}
\right) _{b}.  \tag{19b}
\end{equation}
In this representation the reduced density matrices $\rho _{a}$ and $\rho
_{b}$ read as 
\begin{equation}
\rho _{a}=\left( 
\begin{array}{cc}
c_{+-}^{2} & c_{+-}c_{--} \\ 
c_{+-}c_{--} & c_{-+}^{2}+c_{--}^{2}
\end{array}
\right) ,  \tag{20a}
\end{equation}
and 
\begin{equation}
\rho _{b}=\left( 
\begin{array}{cc}
c_{-+}^{2} & c_{-+}c_{--} \\ 
c_{-+}c_{--} & c_{+-}^{2}+c_{--}^{2}
\end{array}
\right) .  \tag{20b}
\setcounter{equation}{20}
\end{equation}
Note that $\rho _{a}$ and $\rho _{b}$ fulfill the normalization condition $%
\text{Tr}\,\rho _{a}=\text{Tr}\,\rho _{b}=1$. Further, we have that $\det
\rho _{a}=\det \rho _{b}=c_{+-}^{2}c_{-+}^{2}$. Now, according to the
general theory of the Schmidt decomposition [14,15], it follows that the
square of the coefficients $c_{+}$ and $c_{-}$ in Eq.\ (17) are equal to the
eigenvalues $\lambda _{+}$ and $\lambda _{-}$ of $\rho _{a}$ (which are
identical to the two eigenvalues of $\rho _{b}$). These are given by 
\begin{equation}
c_{\pm }^{2}= \lambda _{\pm }= \frac{\text{Tr}\rho _{a}\pm \sqrt{(\text{%
Tr}\rho _{a})^{2}-4\det \rho _{a}}}{2}=\frac{1}{2}\pm \frac{1}{2}\sqrt{%
1-4c_{+-}^{2}c_{-+}^{2}}.  \tag{21}
\end{equation}
From expression (21), it follows immediately that $c_{+}^{2}+c_{-}^{2}=1$.
Moreover, it is important to notice that, whenever we have $%
c_{+-}c_{-+}c_{--}\neq 0$, then necessarily $\lambda _{+}\neq \lambda _{-}$
(i.e., the matrices $\rho _{a}$ and $\rho _{b}$ are nondegenerate). This is
another demonstration of the fact that, when one of the terms in expansion
(2) is missing, then the resulting decomposition in Eq.\ (17) cannot be
maximally entangled. On the other hand, the state vector $\left| S(\phi
_{a})=+1\right\rangle $ ($\left| S(\phi _{a})=-1\right\rangle $) in Eq.\ (17)
is just the eigenvector of $\rho _{a}$ corresponding to the eigenvalue $%
\lambda _{+}$ ($\lambda _{-}$). Similarly, the state vector $\left| S(\phi
_{b})=+1\right\rangle $ ($\left| S(\phi _{b})=-1\right\rangle $) is the
eigenvector of $\rho _{b}$ corresponding to $\lambda _{+}$ ($\lambda _{-}$).

For the particular case in which $c_{+-}^{2}=c_{-+}^{2}=1/\tau ^{2}$ and $%
c_{--}^{2}=1/\tau ^{3}$ (see Eq.\ (12)), we have 
\begin{equation}
c_{\pm }^{2}=\frac{1}{2}\pm \frac{1}{2}\sqrt{1-4\tau ^{-4}}=\frac{1}{2}\pm 
\frac{1}{2}\left( 6\sqrt{5}-13\right) ^{\frac{1}{2}}.  \tag{22}
\end{equation}
So, the approximate values of $c_{+}^{2}$ and $c_{-}^{2}$ giving the maximum
probability $P_{\eta }^{\text{max}}$ are $c_{+}^{2}\simeq 0.822\,\,648$ and $%
c_{-}^{2}\simeq 0.177\,\,352$. Regarding the eigenvectors $\left| S(\phi
_{a})=\pm 1\right\rangle $ and $\left| S(\phi _{b})=\pm 1\right\rangle $
that arise for this particular case, these are given by (supposing for
concreteness that sgn $c_{+-}=$ sgn $c_{-+}=$ sgn $c_{--}$)
\begin{align}
\left| S(\phi _{a})=\pm 1\right\rangle &\equiv \left( 
\begin{array}{c}
f_{\pm } \\ 
g_{\pm }
\end{array}
\right) _{a}  \nonumber \\
&= \left( 
\begin{array}{c}
\frac{2-\sqrt{5}\pm \left( 6\sqrt{5}-13\right) ^{\frac{1}{2}}}{\left[ 12%
\sqrt{5}-26\mp 2\left( 106\sqrt{5}-237\right) ^{\frac{1}{2}}\right] ^{\frac{1%
}{2}}} \\ 
\frac{\left( 10\sqrt{5}-22\right) ^{\frac{1}{2}}}{\left[ 12\sqrt{5}-26\mp
2\left( 106\sqrt{5}-237\right) ^{\frac{1}{2}}\right] ^{\frac{1}{2}}}
\end{array}
\right) _{a},  \tag{23a}
\end{align}
and, (since $\rho _{a}=\rho _{b}$ for this case),$^{3}$%
\begin{equation}
\left| S(\phi _{b})=\pm 1\right\rangle \equiv \left( 
\begin{array}{c}
f_{\pm } \\ 
g_{\pm }
\end{array}
\right) _{b}.  \tag{23b}
\end{equation}
Alternatively, we can also express the eigenvectors $\left| S(\phi _{a})=\pm
1\right\rangle $ as 
\begin{align}
\left| S(\phi _{a})=+1\right\rangle &= \cos \frac{\delta _{a}}{2}\left|
S(\theta _{a})=+1\right\rangle -\sin \frac{\delta _{a}}{2}\left| S(\theta
_{a})=-1\right\rangle ,  \tag{24a} \\
\left| S(\phi _{a})=-1\right\rangle &= \sin \frac{\delta _{a}}{2}\left|
S(\theta _{a})=+1\right\rangle +\cos \frac{\delta _{a}}{2}\left| S(\theta
_{a})=-1\right\rangle ,  \tag{24b}
\end{align}
where $\delta _{a}=\theta _{a}-\phi _{a}$. Thus, for the considered case, we
deduce from Eqs.\ (19a), (23a), and (24a) that $\delta _{a}=2\arccos f_{+}$.
This gives us in principle the two possibilities $\delta _{a}\simeq \pi \pm
68.5414^{\circ }$, and then $\phi _{a}\simeq \theta _{a}\pm 111.4586^{\circ
} $. Similarly, we would find that $\phi _{b}\simeq \theta _{b}\pm
111.4586^{\circ }$. Actually, the specific sign in front of $111.4586^{\circ
}$ to be applied for either $\phi _{a}$ or $\phi _{b}$ does generally depend
on the relative signs of the coefficients $c_{+-}$, $c_{-+}$, and $c_{--}$
in Eq.\ (2).

Finally we note that, as expected, the decomposition (17) fulfilling the
conditions (1a)-(1d) cannot be a product state. To see this, suppose on the
contrary that we have $c_{+}=0$ in Eq.\ (17) so that $\left| \eta
\right\rangle =\left| S(\phi _{a})=-1\right\rangle $\linebreak $\left|
S(\phi _{b})=-1\right\rangle $. For this state, one readily finds that the
fulfillment of conditions (1a)-(1d) is equivalent to requiring,
respectively, that 
\begin{align}
\sin \left( \frac{\delta _{a}}{2}\right) \sin \left( \frac{\delta _{b}}{2}%
\right) &= 0,  \tag{25a} \\
\cos \left( \frac{\delta _{a}}{2}\right) \cos \left( \frac{\delta
_{b}^{^{\prime }}}{2}\right) &= 0,  \tag{25b} \\
\cos \left( \frac{\delta _{a}^{^{\prime }}}{2}\right) \cos \left( \frac{%
\delta _{b}}{2}\right) &= 0,  \tag{25c} \\
\cos \left( \frac{\delta _{a}^{^{\prime }}}{2}\right) \cos \left( \frac{%
\delta _{b}^{^{\prime }}}{2}\right) &\neq 0,  \tag{25d}
\end{align}
where $\delta _{a}=\theta _{a}-\phi _{a}$, $\delta _{b}=\theta _{b}-\phi
_{b} $, $\delta _{a}^{^{\prime }}=\theta _{a}^{^{\prime }}-\phi _{a}$, and $%
\delta _{b}^{^{\prime }}=\theta _{b}^{^{\prime }}-\phi _{b}$. However, as
may be easily checked, the set of conditions (25a)-(25c), on the one hand,
and condition (25d), on the other hand, are mutually incompatible in the
sense that the fulfillment of all three conditions (25a)-(25c) precludes the
fulfillment of condition (25d), and vice versa. Of course a similar remark
applies to the case that $c_{-}=0$. \bigskip

\textbf{Acknowledgment} --- The author wishes to thank Sibasish Ghosh and
G.~Kar for careful reading of an earlier draft of the manuscript and for
valuable comments.
\pagebreak

\begin{center}
{\large NOTES}
\end{center}
\begin{enumerate}
\item  The same type of contradiction is obtained if, in Eqs.\ (1a)-(1d), we
convert all the $-1$'s into $+1$'s, and vice versa. Likewise, the same holds
true if we reverse the sign \textit{only} to the outcomes corresponding to
particle $a$ in each of the Eqs.\ (1a)-(1d). Indeed, the following set of
conditions 
\begin{align*}
P_{\eta }(S(\theta _{a})=-1,S(\theta _{b})=+1) &= 0, \\
P_{\eta }(S(\theta _{a})=+1,S(\theta _{b}^{^{\prime }})=-1) &= 0, \\
P_{\eta }(S(\theta _{a}^{^{\prime }})=+1,S(\theta _{b})=-1) &= 0, \\
P_{\eta }(S(\theta _{a}^{^{\prime }})=+1,S(\theta _{b}^{^{\prime }})=-1)
&> 0,
\end{align*}
does also lead to a Hardy-type contradiction. Naturally, by symmetry, the
same is true if we reverse the sign \textit{only} to the outcomes
corresponding to particle $b$ in each of the Eqs.\ (1a)-(1d).

\item  It is to be noticed that the following relations 
\[
\left| \left\langle S(\theta _{a}^{^{\prime }})=+1\mid S(\theta
_{a})=i\right\rangle \right| =0\Longleftrightarrow \left| \left\langle
S(\theta _{a}^{^{\prime }})=-1\mid S(\theta _{a})=j\right\rangle \right| =0,
\]
and 
\[
\left| \left\langle S(\theta _{b}^{^{\prime }})=+1\mid S(\theta
_{b})=i\right\rangle \right| =0\Longleftrightarrow \left| \left\langle
S(\theta _{b}^{^{\prime }})=-1\mid S(\theta _{b})=j\right\rangle \right| =0,
\]
hold for $i\neq j$.

\item  In terms of the parameter $\tau $, the eigenvectors $\left( 
\begin{array}{c}
f_{\pm } \\ 
g_{\pm }
\end{array}
\right) _{a,b}$ are 
\[
\left( 
\begin{array}{c}
f_{\pm } \\ 
g_{\pm }
\end{array}
\right) _{a,b}=\left( 
\begin{array}{c}
\frac{-\tau ^{-3}\pm \left( 1-4\tau ^{-4}\right) ^{\frac{1}{2}}}{\left[
2-8\tau ^{-4}\mp 2\tau ^{-3}\left( 1-4\tau ^{-4}\right) ^{\frac{1}{2}%
}\right] ^{\frac{1}{2}}} \\ 
\frac{2\tau ^{-\frac{5}{2}}}{\left[ 2-8\tau ^{-4}\mp 2\tau ^{-3}\left(
1-4\tau ^{-4}\right) ^{\frac{1}{2}}\right] ^{\frac{1}{2}}}
\end{array}
\right) _{a,b}.
\]
\pagebreak 
\end{enumerate}

\begin{center}
{\large FIGURE CAPTIONS}{\Large \medskip }
\end{center}

FIG.\ 1 --- Probability of jointly obtaining the results $S(\theta
_{a}+\theta _{1})=-1$ and $S(\theta _{b}+\theta _{2})=-1$, as a function of
the relative angles $\theta _{1}$ and $\theta _{2}$. The surface of
probability remains invariant under a reflection through the vertical planes $%
\theta _{1}=\theta _{2}$, $\theta _{1}+\theta _{2}=2\pi $, $\theta _{1}=\pi $%
, and $\theta _{2}=\pi $. Each of the peaks reaches a maximum value of $%
P_{\eta }^{\text{max}}\simeq 0.090\,\,169$. For the peak near the origin this
maximum is located at the point $\theta _{1}=\theta _{2}\simeq
76.3454^{\circ }$.\bigskip

FIG.\ 2 --- Probability of jointly obtaining the results $S(\theta
_{a}+\theta )=-1$ and $S(\theta _{b}+\theta )=-1$, as a function of the
single parameter $\theta $. This curve corresponds to the interception of
the surface appearing in Fig.\ 1 with the vertical plane $\theta _{1}=\theta _{2}$.
The maximum nonlocal effect is achieved for either $\theta \simeq 76.3454^{\circ
}$ or $\theta \simeq 283.6546^{\circ }$.

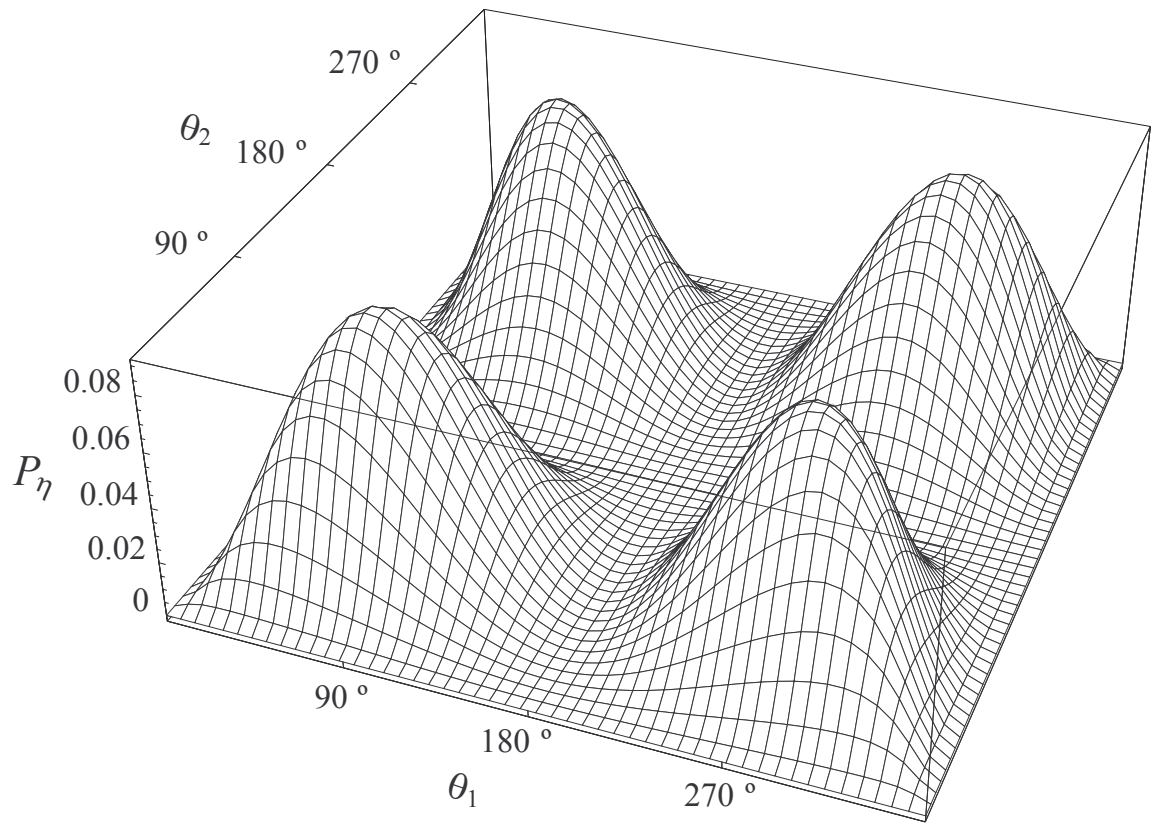
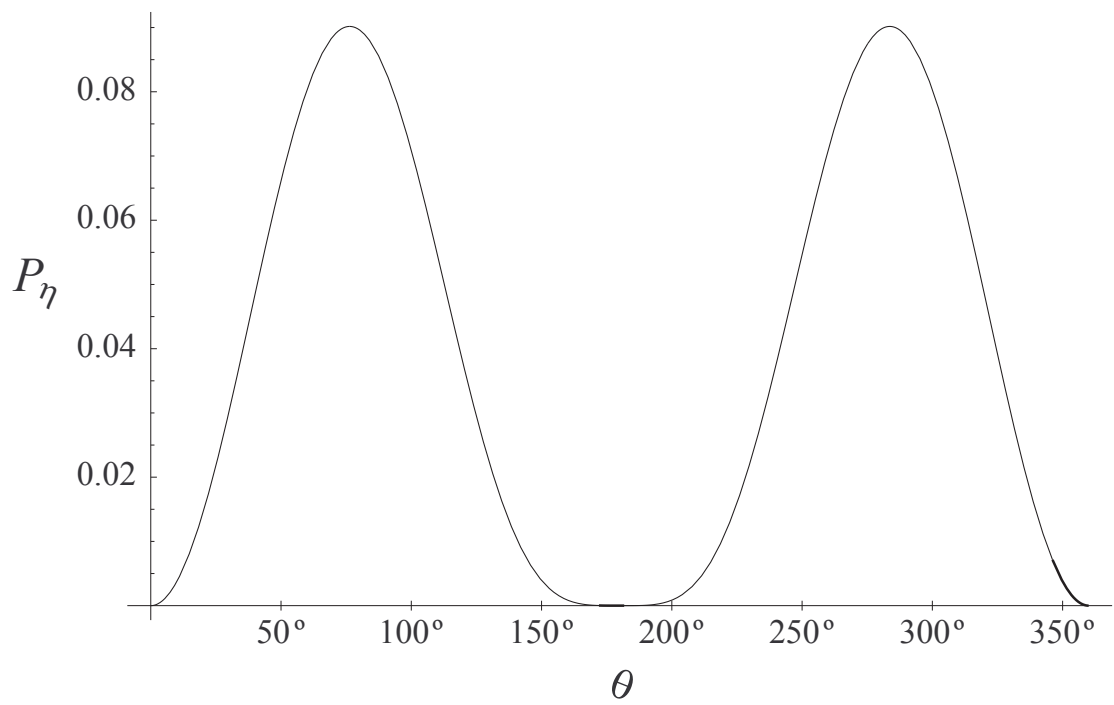
\end{document}